\pdfoutput=1

\documentclass[11pt]{article}

\usepackage[final]{acl}

\usepackage{lipsum}
\usepackage{times}
\usepackage{latexsym}
\usepackage{soul}
\usepackage{url}
\usepackage{amsmath}
\usepackage{amsthm}
\usepackage{booktabs}
\usepackage{algorithm}
\usepackage{algorithmic}
\usepackage{forest}
\usepackage[switch]{lineno}
\usepackage{pifont}
\usepackage{tabularx}     
\usepackage{array}        
\usepackage{multirow}
\usepackage[table]{xcolor}
\usepackage{tabularx}
\usepackage{enumitem}
\usepackage{arydshln}
\hypersetup{    
    urlcolor=darkblue,
}
\usepackage{booktabs}
\usepackage{caption}
\usepackage[T1]{fontenc}

\usepackage[utf8]{inputenc}

\usepackage{microtype}

\usepackage{inconsolata}

\usepackage{graphicx}
\definecolor{cite_color}{HTML}{114083}

\definecolor{url_color}{RGB}{153, 102, 0}

\definecolor{myMintGreen}{HTML}{CEEBD6}

\definecolor{myMintpink}{HTML}{FDE7E9}

\def\eg{\textit{e.g.,} }
\def\ie{\textit{i.e.,} }

\title{Multi-Agent Autonomous Driving Systems with Large Language Models: \\ A Survey of Recent Advances, Resources, and Future Directions}

\author{Yaozu Wu\textsuperscript{1$^*$}, Dongyuan Li\textsuperscript{1$^*$}, Yankai Chen\textsuperscript{2,3$^\dag$}, Renhe Jiang\textsuperscript{1,$\dag$}, \\ \textbf{Henry Peng Zou\textsuperscript{3}}, \textbf{Wei-Chieh Huang\textsuperscript{3}}, \textbf{Yangning Li\textsuperscript{3}}, \\ \textbf{Liancheng Fang\textsuperscript{3}}, \textbf{Zhen Wang}\textsuperscript{1}, \textbf{Philip S. Yu\textsuperscript{3}} 
\\
\\
\textsuperscript{1}The University of Tokyo, 
\textsuperscript{2}Cornell University,
\textsuperscript{3}University of Illinois Chicago
\\
 \small{ 
 {yaozuwu279@gmail.com, lidy@csis.u-tokyo.ac.jp, yankaichen@acm.org, jiangrh@csis.u-tokyo.ac.jp.
 }
 }
}

\begin{document}
\maketitle

\renewcommand{\thefootnote}{}\footnote{$^*$ Equal Contribution. $^\dag$ Corresponding Author.}

\begin{abstract}
\textit{Autonomous Driving Systems (ADSs)} are revolutionizing transportation by reducing human intervention, improving operational efficiency, and enhancing safety.
Large Language Models (LLMs) have been integrated into ADSs to support high-level decision-making through their powerful reasoning, instruction-following, and communication abilities. 
However, LLM-based single-agent ADSs face three major challenges: limited perception, insufficient collaboration, and high computational demands.
To address these issues, recent advances in \textit{LLM-based multi-agent ADSs} leverage language-driven communication and coordination to enhance inter-agent collaboration. 
This paper provides a frontier survey of this emerging intersection between NLP and multi-agent ADSs.
We begin with a background introduction to related concepts, followed by a categorization of existing LLM-based methods based on different agent interaction modes.
We then discuss agent-human interactions in scenarios where LLM-based agents engage with humans.
Finally, we summarize key applications, datasets, and challenges to support future research\footnote{{\url{https://github.com/Yaozuwu/LLM-based_Multi-agent_ADS}}}.
\end{abstract}

\section{Introduction}
\textit{Autonomous driving systems (ADSs)} are redefining driving behaviors, reshaping global transportation networks, and driving a technological revolution~\cite{yurtsever2020survey}.
Traditional ADSs primarily rely on data-driven approaches (as detailed in Appendix~\ref{app:Data-driven}), focusing on system development while overlooking dynamic interactions with the environment.
To enhance engagement with diverse and complex driving scenarios, agentic roles have been incorporated into ADSs~\cite{durante2024agent} using methods like reinforcement learning~\cite{zhang2024multi} and active learning~\cite{lu2024activead}.
Despite notable progress, these methods struggle with ``long-tail'' scenarios, where rare but critical driving situations, such as sudden obstacles, pose significant challenges to model performance.
Furthermore, their ``black-box'' nature limits interpretability, making their decisions difficult to trust.

LLM-based single-agent ADSs help overcome the limitations of data-driven methods~\cite{wang2024survey}. 
{Pre-trained on vast, multi-domain datasets, LLMs excel in knowledge transfer and generalization~\cite{achiam2023gpt}, enabling strong performance in traffic scenarios under zero-shot settings, thus addressing the long-tail issue~\cite{yang2023llm4drive}.} 
{Moreover, techniques such as Reinforcement Learning from Human Feedback (RLHF) and Chain-of-Thought (CoT)~\cite{zhao2023survey}, enhance language-based interaction and logical reasoning, allowing LLMs to make human-like, real-time decisions while providing interpretable and trustworthy feedback across various driving conditions.} 
For instance, Drive-Like-a-Human~\cite{fu2024drive} builds a closed-loop system comprising environment, agent, memory, and expert modules.
The agent interacts with the environment, reflects on expert feedback, and ultimately accumulates experience. For example,  
DiLu~\cite{wen2023dilu} replaces human experts with a reflection module and integrates an LLM-based reasoning engine to enable continuous decision-making. 
Agent-Driver~\cite{mao2024a} designs a tool library to collect environmental data and uses LLMs' cognitive memory and reasoning to improve planning. 

\begin{figure}[t]
\centering
     \includegraphics[width=
     \columnwidth]{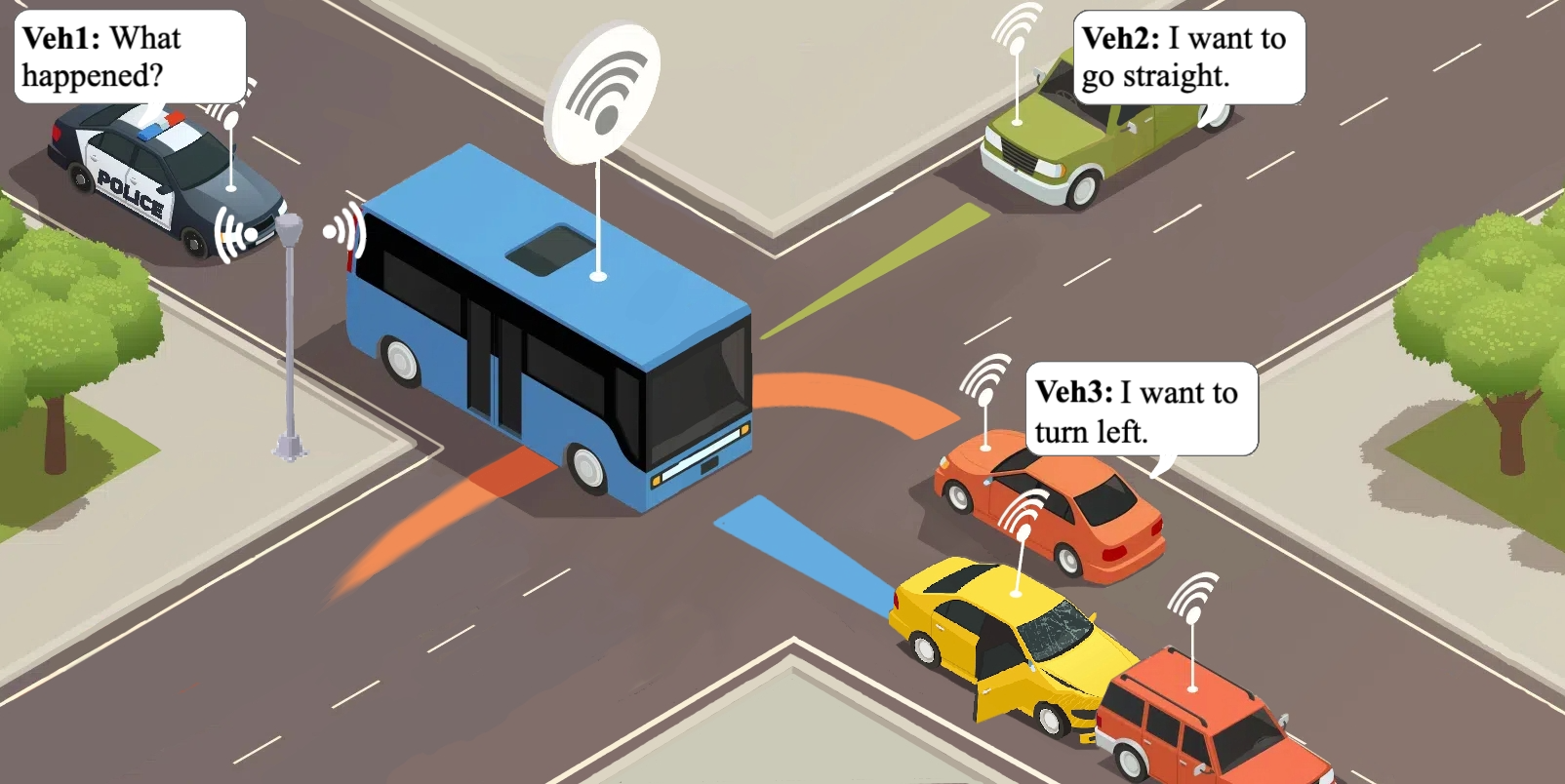}
    \caption{Limitations of LLM-based single-agent ADSs. At an intersection without traffic lights, an accident has occurred ahead, causing Veh1 to be stuck. Due to \textbf{limited perception}, Veh1 is unable to assess the situation and cannot proceed. 
    Veh2 intends to go straight, and Veh3 wants to turn left.
    However, due to \textbf{insufficient collaboration}, they are also unable to navigate the intersection efficiently. 
    Furthermore, due to \textbf{high computing demands}, the lightweight agent on Veh1 struggles to handle the complex driving scenario and has to rely on a more powerful cloud-based agent for assistance.}
    \label{fig:disadvantage}
\end{figure}

\begin{figure*}[t]
\centering
     \includegraphics[width=1\linewidth]{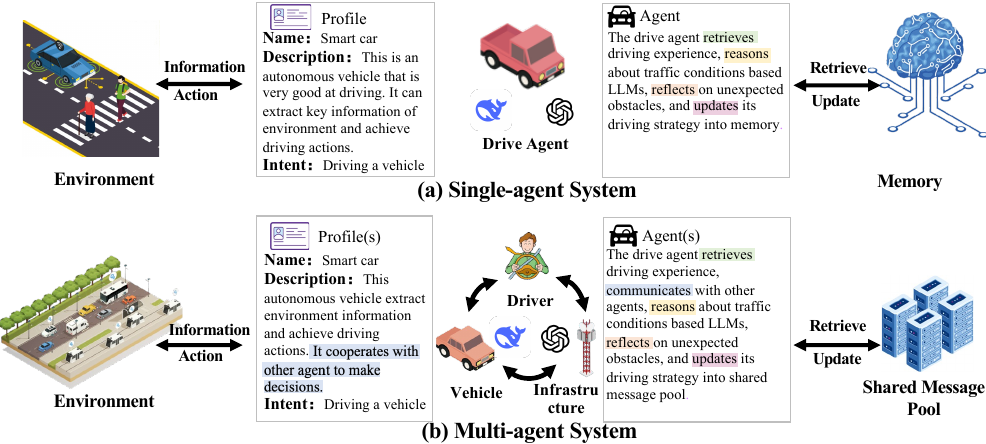}
    \caption{Overview of LLM-based (a) single- and (b) multi-agent ADSs, with key terms and differences highlighted.
    }
    \label{fig:MAS}
\end{figure*}
However, as shown in Figure~\ref{fig:disadvantage}, researchers have identified three critical limitations of LLM-based single-agent ADSs in complex traffic environments: 
\ding{182}~\textbf{Limited Perception: } 
LLMs can only respond to sensor inputs and lack predictive and generalization capabilities. As a result, LLM-based single-agent ADSs cannot complement incomplete sensor information and thus miss critical information in driving scenarios, such as pedestrians or vehicles hidden in complex intersection environments~\cite{hu2024collaborative}.
\ding{183}~\textbf{Insufficient Collaboration: } A single LLM-based agent cannot coordinate with other vehicles or infrastructure, leading to suboptimal performance in scenarios requiring multi-agent interactions, such as merging of lanes or navigate roundabouts~\cite{malik2021collaborative}. 
\ding{184}~\textbf{High Computational Demands: } With billions of parameters in LLMs, these methods require substantial independent computational resources, making real-time deployment challenging, particularly in resource-limited in-vehicle systems~\cite{cui2024personalizedautonomousdrivinglarge}. 

To address these limitations, LLM-based multi-agent ADSs enable distinct agents to communicate and collaborate, improving safety and performance. 
{First}, LLMs enhance {contextual awareness} by allowing agents to share data, extend their perceptual range, and enhance the detection of occluded objects in complex environments~\cite{hu2024collaborative}. 
{Second}, real-time coordination among LLM-based agents mitigates {insufficient collaboration}, enabling joint decisions in tasks like lane merging and roundabout navigation, ultimately leading to safer and more efficient driving operations~\cite{malik2021collaborative}. 
{Third}, LLMs optimize {computational efficiency} by distributing tasks across agents, reducing individual load and enabling real-time processing in resource-limited systems.

As LLM capabilities continue to advance, they are playing an increasingly significant role in ADS as intelligent driving assistants.
Several reviews have focused on two primary aspects: 
\textit{\textbf{i)}} the integration of LLMs in data-driven methods~\cite{yang2023llm4drive,li2023towards} and \textit{\textbf{ii)}} the applications of specific LLM types, such as vision-based~\cite{zhou2024vision} and multimodal-based~\cite{fourati2024xlm,cui2024survey} models in ADS.
However, no comprehensive survey has systematically examined the emerging field of LLM-based multi-agent ADSs.
This gap motivates us to provide a comprehensive review that consolidates existing knowledge and offers insights to guide future research and the development of advanced ADSs.

In this study, we present a comprehensive survey of LLM-based multi-agent systems. Specifically, Section~\ref{sec:MAS} introduces the core concepts, 
including \textit{agent environments and profiles}, \textit{inter-agent interaction mechanisms}, and \textit{agent-human interactions}. 
Section~\ref{sec:agent} provides a structured review of 
existing studies: 
\textit{multi-vehicle interaction}, \textit{vehicle-infrastructure interaction}, and \textit{vehicle-assistant interaction}.
As agent capabilities continue to grow, human-vehicle co-driving is emerging as the dominant autonomous driving paradigm, with human playing an increasingly vital role. 
Humans collaborate with agents by providing guidance or supervising their behavior. Therefore, we consider humans as special virtual agents and examine human-agent interactions in Section~\ref{sec:human-vehicle}.
Section~\ref{sec:application} explores various applications, while Section~\ref{sec:dataset} compiles a comprehensive collection of public datasets and open-source resources.
Section~\ref{sec:opportunities} discusses existing challenges and future research directions. Finally, Section~\ref{sec:conclusion} concludes the study.

\section{LLM-based Agents for ADS}
\label{sec:MAS}
\subsection{LLM-based Single-Agent ADS}
Achieving human-level driving is an ultimate goal of ADS. As shown in Figure~\ref{fig:MAS}(a), the LLM-based single agent retrieves past driving experiences from the memory, integrates them with real-time environmental information for reasoning, and makes driving decisions. Additionally, the driving agent reflects on its decision and updates its memory accordingly, ensuring safe and efficient driving actions. 
However, the complex and dynamic nature of real-world driving scenarios, where interactions with other vehicles significantly impact decision-making, suggests that neglecting these interactions can lead to suboptimal or unsafe driving outcomes.

\subsection{LLM-based Multi-Agent ADS}
With interactions among multiple agents, LLM-based multi-agent ADS leverages collective intelligence and specialized skills, with each agent playing a distinct role, communicating and collaborating within the system. This enhances the efficiency and safety of autonomous driving. 
Below, we introduce the LLM-based multi-agent ADS, as shown in Figure~\ref{fig:MAS}(b), and provide a detailed analysis of its three key modules: {Agent Environment and Profile}, {LLM-based Multi-Agent Interaction}, and {LLM-based Agent-Human Interaction}. 

\subsubsection{Agent Environment and Profile}
Similar to the single-agent architecture in Figure~\ref{fig:MAS}(a), multi-agent systems first obtain relevant information from their \textit{environments}, enabling them to make informed decisions and take appropriate actions.
The environmental conditions define the settings and necessary context for agents in LLM-based multi-agent ADS to operate effectively.
Generally, there are two environment types, \ie {physical environment} and {simulation environment}.

\definecolor{tblHeader}{RGB}{217,225,242}

\definecolor{rowA}{RGB}{245,250,255}
\definecolor{rowB}{RGB}{245,255,245}
\definecolor{rowC}{RGB}{255,245,240}

\definecolor{methodBlue}{RGB}{51,102,153}   
\definecolor{methodGreen}{RGB}{60,150, 90}  
\definecolor{methodRed}{RGB}{153, 51, 51}   

\colorlet{rowAtext}{methodBlue}
\colorlet{rowBtext}{methodGreen}
\colorlet{rowCtext}{methodRed}

\begin{table}[t]
  \caption{Comparison of Agent Profiling Methods.}
  \centering
  \tiny                             
  \setlength{\tabcolsep}{5pt}       

  \begin{tabularx}{\columnwidth}{
      p{1.5cm}
      >{\raggedright\arraybackslash}X
      >{\raggedright\arraybackslash}X
    }
    \toprule
    \rowcolor{tblHeader}
    \textbf{Method} &
    \textbf{Advantage} &
    \textbf{Limitation} \\
    \midrule

    \rowcolor{rowA}
    \textbf{\textcolor{rowAtext}{\textit{Pre-defined}}} &
    Rely on prior knowledge to \textbf{reduce the difficulty of scenario modeling} and \textbf{embed strict safety rules and regulatory constraints}. &
    \textbf{Labor-intensive} to create and maintain, and \textbf{lacks adaptability} to novel or dynamic autonomous driving scenarios. \\
    \addlinespace[3pt]

    \rowcolor{rowB}
    \textbf{\textcolor{rowBtext}{\textit{Model-generated}}} &
    Synthesize new agent roles on-the-fly, letting simulators or fleets \textbf{adapt to unseen driving contexts}. &
    Generated profiles \textbf{may violate traffic laws} and \textbf{have limited understanding of safety-critical environments}. \\
    \addlinespace[3pt]

    \rowcolor{rowC}
    \textbf{\textcolor{rowCtext}{\textit{Data-derived }}} &
    Can \textbf{learn complex, real-world driving behaviors and patterns} from large datasets, potentially improving naturalistic interactions. &
    Coverage remains \textbf{limited by the availability of vast, high-quality autonomous driving data}, and \textbf{privacy or commercial constraints} may restrict data sharing. \\

    \bottomrule
  \end{tabularx}
  \label{tab:agent-profiles}
\end{table}

\textbf{\textit{Physical environment}} represents the real-world setting where driver agents gather information using various sensors, such as cameras and LiDAR, and interact with other traffic participants.
However, due to the high cost of vehicles and strict regulations on public roads, collecting large amounts of data in real world is impractical. As a viable alternative, the \textbf{\textit{Simulation environment}} provides a simulated setting constructed by humans.
It can accurately model specific conditions without incurring the high costs and complexities associated with real-world data collection, allowing agents to freely test actions and strategies across a variety of scenarios~\cite{dosovitskiy2017carla}.

In LLM-based multi-agent systems, each agent is assigned distinct roles with specific functions through \textit{profiles}, enabling them to collaborate on complex driving tasks or simulate intricate traffic scenarios.
These profiles are crucial in defining the functionality of the agent, its interaction with the environment, and its collaboration with other agents.
Existing work~\cite{li2024survey} generates agent profiles using three types of methods: {Pre-defined}, {Model-generated}, and {Data-derived}. 

Table~\ref{tab:agent-profiles} summarizes the advantages and limitations of different agent profiling methods in ADSs. Specifically, 
within \textbf{\textit{Pre-defined methods}}, system designers explicitly define agent profiles based on prior knowledge and the analysis of complex scenarios~\cite{chen2024edge}.
Each agent has unique attributes and behavior patterns that can be adjusted based on the scenario. 
In driving environments, the objectives of ADS require the collaboration of vehicle agents, infrastructure agents, and drivers. In particular, 
\ding{182}~Vehicle agents denote various types of autonomous vehicles, traveling according to preset routes and traffic rules, while communicating and collaborating with other vehicles and driver agents.
\ding{183}~Infrastructure agents, \eg traffic lights, road condition monitors, and parking facilities, provide real-time traffic information and instructions, influencing the behavior of driver and vehicle agents. However, manually crafting such roles is labor-intensive and often brittle when scenarios shift, which has stimulated interest in automatic profile construction, either generated by LLMs or extracted from large-scale datasets. \textbf{\textit{Model-generated methods}} create agent profiles using advanced LLMs based on the interaction context and the goals that need to be accomplished~\cite{zhou2024algpt} and \textbf{\textit{Data-derived Profile}} design agent profiles based on pre-existing datasets~\cite{guo2024large}.

\subsubsection{LLM-based Multi-Agent Interaction}
In LLM-based multi-agent ADS, effective 
communication and coordination among agents are crucial to improve collective intelligence and solve complex traffic scenarios. 
Agent interactions depend on both the interaction mode and the underlying interaction structure, as summarized in Table~\ref{tab:driving_dimensions}.

\textit{\textbf{The interaction mode}} can be classified as: \textit{cooperative}, \textit{competitive}, and \textit{debate} mode. 
\ding{182}~In cooperative mode, agents work together to achieve shared objectives by exchanging information~\cite{DBLP:conf/naacl/ChenZH24,jin2024surrealdriver}.
\ding{183}~In competitive mode, agents strive to accomplish their individual goals and compete with others~\cite{yao2024comal}.
\ding{184}~The Debate mode enables agents to debate with each other, propose their own solutions, criticize the solutions of other agents, and collaboratively identify optimal strategies~\cite{DBLP:conf/emnlp/Liang0JW00Y0T24}.

\begin{figure}[t]
\centering
     \includegraphics[width=\columnwidth]{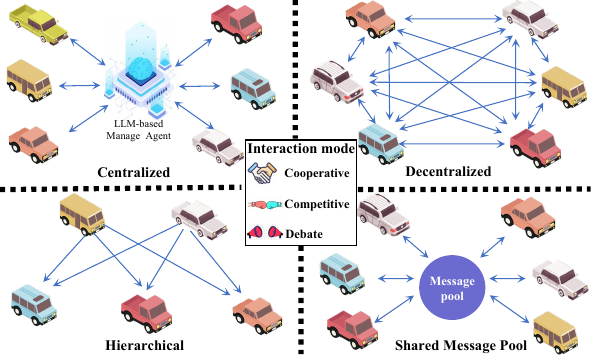}
    \caption{Different interaction modes and structures.}
    \label{fig:structure}
\end{figure}

\textit{\textbf{The interaction structure}} delineates the architecture of communication networks within LLM-based multi-agent ADS, including \textit{centralized}, \textit{decentralized}, \textit{hierarchical}, and \textit{shared message pool} structures, as shown in Figure~\ref{fig:structure}.
Specifically, \ding{182}~the centralized interaction structures defines a central agent or a group of central agents to manage interactions among all agents~\cite{zhou2024algpt}.
\ding{183}~The decentralized interaction structure allows for direct communication between agents, with all agents being equal to each other~\cite{hu2024agentscomerge}.
\ding{184}~Hierarchical structures focus on interactions within a layer or with adjacent layers~\cite{DBLP:conf/coling/OhmerDB22}.
\ding{185}~The shared memory interaction structure maintains a shared message pool, allowing agents to send and extract the necessary information~\cite{jiang2024koma}.
We provide a more detailed introduction to LLM-based multi-agent ADSs based on their interaction structures and modes in Section~\ref{sec:agent}. 

\subsubsection{LLM-based Agent-Human Interaction}

Recent studies show that human-machine co-driving systems use LLMs to improve agent-human interactions, enabling vehicles to communicate and collaborate seamlessly with human drivers through natural language~\cite{DBLP:conf/emnlp/FengCQLC0W24, zou2025llmbasedhumanagentcollaborationinteraction, zou2025call}.
This allows vehicles to better understand and respond to human intent, provide context-aware responses, enhance driving safety and comfort, and offer personalized recommendations based on driver preferences.
Humans also play a crucial role in guiding and supervising agent behavior, enhancing the agents' capabilities while ensuring safety. 
We examine the role of humans as special virtual agents and explore agent-human interaction dynamics in Section~\ref{sec:human-vehicle}.

\section{LLM-based Multi-Agent Interaction}
\label{sec:agent}

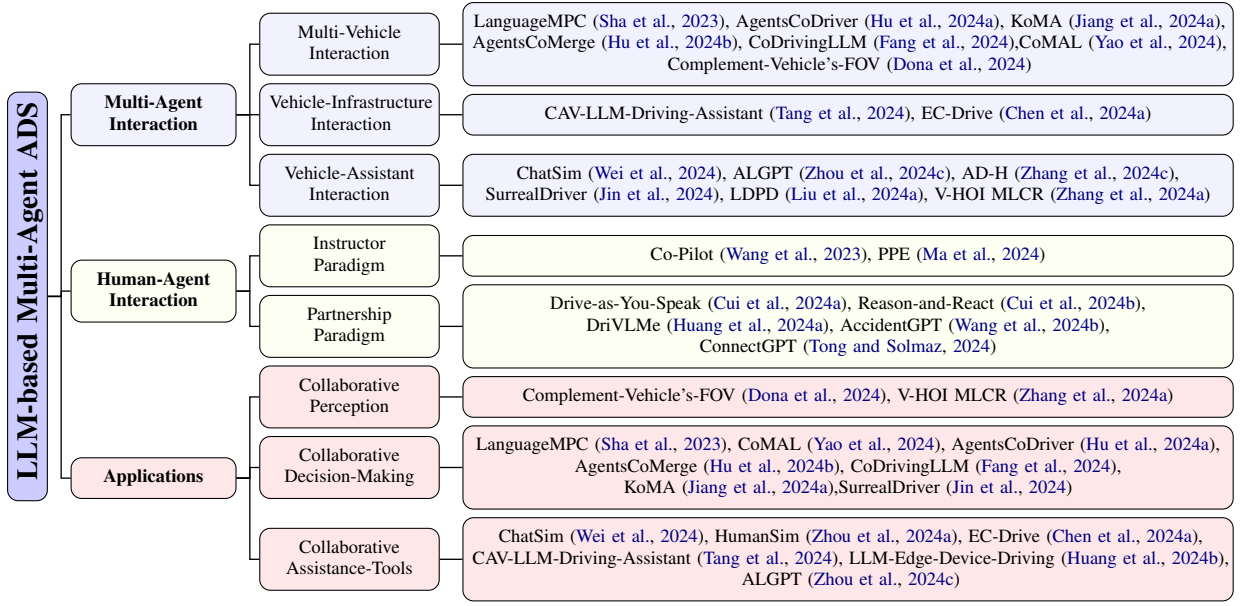
\begin{figure*}[t]
\centering
\begin{forest}
for tree={
    font=\footnotesize, 
    grow=east, 
    draw,
    rounded corners,
    text width=6em, 
    anchor=center,
    align=center, 
    text centered,
    parent anchor=east,
    child anchor=west,
    edge path={
        \noexpand\path[\forestoption{edge}, line width=0.5pt]
        (!u.parent anchor) -- ++(5pt,0) |- (.child anchor)\forestoption{edge label};
    },
    s sep=1mm, 
    l sep=3mm, 
}
[
\rotatebox{90}{LLM-based Multi-Agent ADS} , fill=blue!20, font=\bfseries, text width=0.7em, anchor=center, align=center, 
    [
    Applications, fill=myMintpink, font=\scriptsize\bfseries, text width=5em
        [
        Collaborative \\ Assistance-Tools, fill=myMintpink, font=\scriptsize, text width=5.5em
            [{ChatSim~\cite{wei2024editable}, HumanSim~\cite{zhou2024humansim}, EC-Drive~\cite{chen2024edge}, \\ CAV-LLM-Driving-Assistant~\cite{tang2024test}, LLM-Edge-Device-Driving~\cite{huang2024efficient}, \\ ALGPT~\cite{zhou2024algpt}}, fill=myMintpink, font=\scriptsize, text width=25.8em]
        ]
        [
        Collaborative \\ Decision-Making, fill=myMintpink, font=\scriptsize, text width=5.5em
            [{LanguageMPC~\cite{sha2023languagempc}, CoMAL~\cite{yao2024comal}, AgentsCoDriver~\cite{hu2024agentscodriver}, \\ AgentsCoMerge~\cite{hu2024agentscomerge}, CoDrivingLLM~\cite{fang2024towards}, \\ KoMA~\cite{jiang2024koma},SurrealDriver~\cite{jin2024surrealdriver}}, fill=myMintpink, font=\scriptsize, text width=25.8em]
        ]
        [
        Collaborative \\ Perception, fill=myMintpink, font=\scriptsize, text width=5.5em
            [{Complement-Vehicle's-FOV~\cite{dona2024tapping}, V-HOI MLCR~\cite{zhang2024enhancing}}, fill=myMintpink, font=\scriptsize, text width=25.8em]
        ]
    ]
    [
    Human-Agent \\ Interaction, fill=lime!5, font=\scriptsize\bfseries, text width=5em
        [
        Partnership \\ Paradigm, fill=lime!5, font=\scriptsize, text width=5.5em
            [{Drive-as-You-Speak~\cite{cui2024drive}, Reason-and-React~\cite{cui2024receive}, \\ DriVLMe~\cite{huang2024drivlme}, AccidentGPT~\cite{wang2024accidentgpt}, \\ ConnectGPT~\cite{tong2024connectgpt}}, fill=lime!5, font=\scriptsize, text width=25.8em
            ]
        ]    
        [
        Instructor \\ Paradigm, fill=lime!5, font=\scriptsize, text width=5.5em
        [{Co-Pilot~\cite{wang2023chatgpt}, PPE~\cite{ma2024learning}}, , fill=lime!5, font=\scriptsize, text width=25.8em
            ]
        ]
    ]
    [
    Multi-Agent \\ Interaction, fill=blue!5, font=\scriptsize\bfseries, text width=5em
        [
        Vehicle-Assistant \\ Interaction, fill=blue!5, font=\scriptsize, text width=5.5em
            [{ChatSim~\cite{wei2024editable}, ALGPT~\cite{zhou2024algpt}, AD-H~\cite{zhang2024ad}, \\ SurrealDriver~\cite{jin2024surrealdriver}, LDPD~\cite{liu2024language}, V-HOI MLCR~\cite{zhang2024enhancing}}, fill=blue!5, font=\scriptsize, text width=25.8em]
        ]
        [
        Vehicle-Infrastructure \\ Interaction, fill=blue!5, font=\scriptsize, text width=5.5em
            [{CAV-LLM-Driving-Assistant~\cite{tang2024test}, EC-Drive~\cite{chen2024edge}}, fill=blue!5, font=\scriptsize, text width=25.8em]
        ]
        [
        Multi-Vehicle \\ Interaction, fill=blue!5, font=\scriptsize, text width=5.5em
            [{LanguageMPC~\cite{sha2023languagempc}, AgentsCoDriver~\cite{hu2024agentscodriver}, KoMA~\cite{jiang2024koma}, \\AgentsCoMerge~\cite{hu2024agentscomerge}, CoDrivingLLM~\cite{fang2024towards},CoMAL~\cite{yao2024comal}, \\ Complement-Vehicle's-FOV~\cite{dona2024tapping}}, fill=blue!5, font=\scriptsize, text width=25.8em]
        ]
    ]
]
\end{forest}
\caption{A taxonomy of LLM-based Multi-Agent Autonomous Driving Systems.}
\label{fig:taxonomy}
\end{figure*}

Mutual interaction is central to multi-agent ADSs, enabling systems to solve complex problems beyond the capabilities of a single agent. 
Through information exchange and coordinated decision-making, multiple agents effectively complete shared tasks and achieve overarching objectives~\cite{li2024survey}. 
This section reviews recent studies on multi-agent ADSs, emphasizing interactions among vehicles, infrastructures, and assisted agents in driving scenarios. 
As shown in Figure~\ref{fig:taxonomy}, we categorize existing methods into three interaction types: \textit{multi-vehicle interaction}, \textit{vehicle-infrastructure interaction}, and \textit{vehicle-assistant interaction}.

\subsection{Multi-Vehicle Interaction}
Multi-vehicle interactions involve multiple autonomous vehicles powered by LLMs exchanging real-time information, such as locations, speeds, sensor data, and intended trajectories.
By sharing partial observations of the environment or negotiating maneuvers, multiple vehicles overcome the inherent limitations of single-agent ADS, such as restricted perception and lack of collaboration. 

Typically, these interactions operate in a cooperative mode with varying architectures.
LanguageMPC~\cite{sha2023languagempc} employs a centralized structure, where a central agent acts as the fleet's "brain," providing optimized coordination and 
control commands to each vehicle agent.
In contrast, other decentralized methods~\cite{fang2024towards,dona2024tapping} 
treat all agents as peers, allowing direct vehicle-to-vehicle communication without central bottlenecks.
For instance, AgentsCoDriver~\cite{hu2024agentscodriver} designs an adaptive communication module that generates context-aware messages for inter-agent 
communication when the agent deems it necessary.
AgentsCoMerge~\cite{hu2024agentscomerge} and CoDrivingLLM~\cite{fang2024towards} incorporate agent communication directly into the reasoning process, facilitating real-time intention sharing and proactive negotiation before decision-making.
Additionally, KoMA~\cite{jiang2024koma} and CoMAL~\cite{yao2024comal} build a distributed shared memory pool, allowing agents to send and retrieve the necessary information to facilitate scalable interaction between agents.

\subsection{Vehicle-Infrastructure Interaction}
The interaction between vehicles and external agents, such as traffic lights, roadside sensors, and LLM-powered control centers, not only helps autonomous vehicles make more intelligent decisions but also alleviates on-board computing requirements.
This enables LLM-based multi-agent ADSs to operate effectively in real-world environments.
EC-Drive~\cite{chen2024edge} proposes an Edge-Cloud collaboration framework with a hierarchical interaction structure.
The edge agent processes real-time sensor data and makes preliminary decisions under normal conditions.
When anomalies are detected or the edge agent generates a low-confidence prediction, the system flags these instances and uploads them to the cloud agent equipped with LLMs. 
The cloud agent then performs detailed reasoning to generate optimized decisions and combines them with the output of the edge agent to update the driving plan.
Following a similar architecture, \citet{tang2024test} uses agents deployed on remote clouds or network edges to assist connected driving agents in handling complex driving decisions.

\subsection{Vehicle-Assistant Interaction}
Beyond the interactions between the primary agents in driving scenarios, 
additional interactions among assisted agents play a crucial role in LLM-based multiagent ADSs.
Both ChatSim~\cite{wei2024editable} and ALGPT~\cite{zhou2024algpt} employ a manager (PM) agent to interpret user instructions and coordinate tasks among other agents. 
ChatSim~\cite{wei2024editable} adopts a centralized structure in which the PM agent decouples an overall demand into specific subtasks and dispatches instructions to other team agents.  
Similarly, the PM agent in ALGPT~\cite{zhou2024algpt} formulates a work plan upon receiving user commands and assembles an agent team with the plan. 
Specifically, agents no longer communicate point-to-point with each other but instead communicate through a shared message pool, greatly improving efficiency. 

Additionally, hierarchical agent architectures further enhance the performance and effectiveness of LLM-based multi-agent ADSs.
AD-H~\cite{zhang2024ad} assigns high-level reasoning tasks to the multimodal LLM-based planner agent while delegating low-level control signal generation to a lightweight controller agent.
These agents interact through mid-level commands generated by the multimodal LLMs.
In LDPD~\cite{liu2024language}, the teacher agent leverages the LLM for complex cooperative decision reasoning and trains smaller student agents via its own decision demonstrations to achieve cooperative decision-making.
SurrealDriver~\cite{jin2024surrealdriver} introduces a CoachAgent to evaluate DriverAgent's driving behavior and provide guidelines for continuous improvement.

Different from the conventional collaborative interaction mode, V-HOI~\cite{zhang2024enhancing} proposes a hybrid interaction mode that blends collaboration with debate.
It establishes various agents across different LLMs to evaluate reasoning logic from different aspects, enabling cross-agent reasoning. 
This process culminates in a debate-style integration of responses from various LLMs, improving predictions for enhanced decision-making.

\section{LLM-based Agent-Human Interaction}
\label{sec:human-vehicle}
Depending on the roles of human assume when interacting with agents, we classify current methods as: \textit{instructor paradigm} and \textit{partnership paradigm}.

\subsection{Instructor Paradigm}
In Figure~\ref{fig:human-agent}, the instructor paradigm involves agents interacting with humans in a conversational manner, where humans act as ``tutors'' to offer quantitative and qualitative feedback to improve agent decision-making~\cite{li2017dialogue}.
Quantitative feedback typically includes binary evaluations or ratings, while qualitative feedback consists of language suggestions for refinement. 
Agents incorporate this feedback to adapt and perform better in complex driving scenarios.
For instance, \citet{wang2023chatgpt} propose ``Expert-Oriented Black-box Tuning'', where domain experts provide feedback to optimize model performance. 
Similarly, \citet{ma2024learning} present a human-guided learning pipeline that integrates driver feedback to refine agent decision-making.

\subsection{Partnership Paradigm}
\begin{figure}[t]
\centering
    \includegraphics[width=1.05\columnwidth]{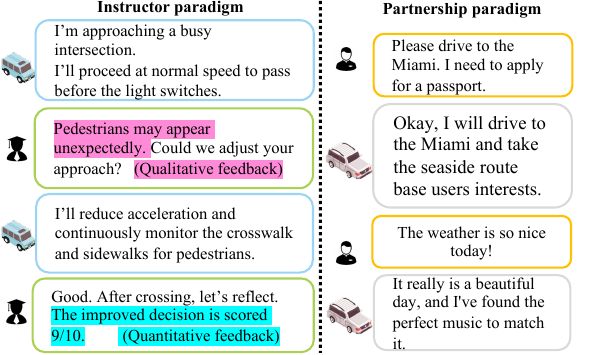}
    \caption{Two modes of agent-human interaction.}
    \label{fig:human-agent}
\end{figure}

As shown in Figure~\ref{fig:human-agent}, the partnership paradigm emphasizes collaboration, where agents and humans interact as equals to accomplish complex driving tasks. 
In this paradigm, agents assist in decision-making by adapting to individual driver preferences and real-time traffic conditions.
For instance, Talk2Drive~\cite{cui2024personalizedautonomousdrivinglarge}, 
DaYS~\cite{cui2024drive} and Receive~\cite{cui2024receive} utilize memory modules to store human-vehicle interactions, enabling a more personalized driving experience based on individual driver preferences, such as overtaking speed and following distance. 
Additionally, infrastructure agents in AccidentGPT~\cite{wang2024accidentgpt} and ConnectGPT~\cite{tong2024connectgpt} connect vehicles to monitor traffic conditions, identify potential hazards, and provide proactive safety warnings, blind spot alerts, and driving suggestions through agent-human interaction.

\section{Applications}
\label{sec:application}
\subsection{Collaborative Perception}
Despite significant advancements in the perception modules of ADS, LLM-based single-agent ADS continues to face substantial challenges, including constrained sensing ranges and persistent occlusion issues~\cite{han2023collaborative}. 
These two key limitations hinder their comprehensive understanding of the driving environment and can lead to suboptimal decision-making, especially in complex and dynamic traffic scenarios~\cite{hu2024collaborative}.

\citet{dona2024tapping} propose a multi-agent cooperative framework that enhances the ego vehicle's field-of-view (FOV) by integrating 
complementary visual perspectives through inter-vehicle dialogues mediated by onboard LLMs, significantly expanding the ego vehicle's environmental comprehension. 
However, in complex road scenarios, reliance on a single LLM can lead to erroneous interpretations and hallucinatory predictions when processing complex traffic situations. 
To address this limitation, V-HOI MLCR~\cite{zhang2024enhancing} introduces a collaborative debate framework among different LLMs for video-based Human-Object Interaction (HOI) detection tasks.
This framework first implements a Cross-Agent Reasoning scheme, assigning distinct roles to various agents within an LLM to conduct reasoning from multiple perspectives.
Subsequently, a cyclic debate mechanism is employed to evaluate and aggregate responses from multiple agents, culminating in the final outcome.

\subsection{Collaborative Decision-Making}
After obtaining environmental information, the ADS performs three core functions: route planning, trajectory optimization, and real-time decision-making.
In complex traffic scenarios such as roundabout navigation and lane merging, LLM-based multi-agent systems enable coordinated motion planning through three key mechanisms: \ding{182}~real-time intention sharing between agents, \ding{183}~adaptive communication protocols, and \ding{184}~dynamic negotiation frameworks. 
This collaborative architecture allows ADS to precisely coordinate their trajectories, maneuver strategies, and environmental interactions while maintaining operational safety.

LanguageMPC~\cite{sha2023languagempc} uses LLMs to perform scenario analysis and decision-making.
Additionally, it introduces a multi-vehicle control method where distributed LLMs govern individual vehicle operations, while a central LLM facilitates multi-vehicle communication and coordination.
AgentsCoDriver~\cite{hu2024agentscodriver} presents a comprehensive LLM-based multi-vehicle collaborative decision-making framework with life-long learning capabilities, moving the field towards practical applications. 
This framework consists of five parts, as follows: the observation module, cognitive memory module, and reasoning engine support the high-level decision-making process for AD; the communication module enables negotiation and collaboration among vehicles; and the reinforcement reflection module reflects the output and decision-making process. 
Similarly, AgentsCoMerge~\cite{hu2024agentscomerge} combines vision-based and text-based scene understanding to gather essential environmental information and incorporates a hierarchical planning module to allow agents to make informed decisions and effectively plan trajectories. 
Instead of directly interacting with each other, agents in KoMA~\cite{jiang2024koma} analyze and infer the intentions of surrounding vehicles via an interaction module to enhance decision-making.
It also introduces a shared memory module to store successful driving experiences and a ranking-based reflection module to review them.

\subsection{Collaborative Cloud-Edge Deployment}
Although many innovative studies have explored the application of LLM-based multi-agent ADS, significant technical challenges remain in deploying LLMs locally on autonomous vehicles due to their huge computational resource requirements~\cite{DBLP:conf/emnlp/0015WCWFWWZY24}.
To address these issues, \citet{tang2024test} apply remote LLMs to provide assistance for connected autonomous vehicles, which communicate between themselves and with LLMs via vehicle-to-everything technologies.
Moreover, this study evaluates LLMs' comprehension of driving theory and skills in a manner akin to human driver tests.
However, remote LLM deployment can introduce inference latency, posing risks in emergency scenarios.
To further improve system efficiency, \citet{chen2024edge} introduce a novel edge-cloud collaborative ADS with drift detection capabilities, using small LLMs on edge devices and GPT-4 on cloud to process motion planning data and complex inference tasks, respectively.

\begin{table*}[t]
\scriptsize
\centering
\caption{Single-agent and multi-agent autonomous driving datasets. }
\resizebox{1\linewidth}{!}{%
\begin{tabular}{llccc}
\toprule
\textbf{Datasets} & \textbf{Dataset Type} & \textbf{Sensor Type}            & \textbf{Tasks}                                                    \\ \hline
KITTI~\cite{geiger2012we}      & Single-agent       & Camera, LiDAR            & 2D/3D detection, tracking, depth estimation                       \\
nuScenes~\cite{caesar2020nuscenes}    & Single-agent      & Cameras, LiDAR, Radars & 3D detection, tracking, trajectory forecasting                    \\
BDD100K~\cite{yu2020bdd100k}   & Single-agent        & Camera                          & Object detection, lane detection, segmentation \\
Waymo~\cite{sun2020scalability}   & Single-agent          & Camera, LiDAR, Radars  & 2D/3D detection, tracking, domain adaptation                      \\ 
BDD-X~\cite{kim2018textual}    & Single-agent         & BDD                             & Object detection, driving scenario captioning                     \\
nuScenes-QA~\cite{qian2024nuscenes}  & Single-agent     & nuScenes                        & 3D detection, tracking, visual QA                                 \\
DriveLM~\cite{sima2025drivelm}      & Single-agent     & nuScenes, Waymo                        & Multi-modal planning, question answering                          \\ 
DAIR-V2X~\cite{yu2022dair}  & Multi-agent       & Camera, LiDAR (multi-vehicle)   & Cooperative perception, tracking                                  \\
TUMTraf-V2X~\cite{zimmer2024tumtraf}     & Multi-agent   & Multi-vehicle camera, LiDAR     & Cooperative perception, multi-agent tracking                      \\
V2V4Real~\cite{xu2023v2v4real}    & Multi-agent      & Multi-vehicle camera, LiDAR     & Cooperative detection, tracking                                   \\
V2XSet~\cite{xu2022v2x}      & Multi-agent       & Multi-vehicle camera, LiDAR     & Multi-agent detection, tracking                                   \\ \bottomrule
\end{tabular}
}
\label{Table-dataset}
\end{table*}

\subsection{Collaborative Assistance-Tools}
The long-term data accumulation in both industry and academia has enabled great success in highway driving and automatic parking~\cite{liu2024survey}. 
However, collecting real-world data remains costly, especially for multi-agents or customized scenarios. 
Additionally, the uncontrollable nature of real scenarios makes it challenging to capture certain corner cases.
To address these issues, many LLM-based studies focus on simulating multi-agent ADS, offering a cost-effective alternative to real-world data collection.
For example, ChatSim~\cite{wei2024editable} provides editable photo-realistic 3D driving scenario simulations via natural language commands and external digital assets. 
The system leverages multiple LLM agents with specialized roles to decompose complex commands into specific editing tasks, introducing novel McNeRF and Mclight methods that generate customized high-quality output.
HumanSim~\cite{zhou2024humansim} integrates LLMs to simulate human-like driving behaviors in multi-agent systems via pre-defined driver characters. 
By employing navigation strategies, HumanSim facilitates behavior-level control of vehicle movements, making it easier to generate corner cases in multi-agent environments. In addition, ALGPT~\cite{zhou2024algpt} uses a multiagent cooperative framework for open-vocabulary, multimodal auto-annotation in autonomous driving. It introduces a Standard Operating Procedure to define agent roles and share documentation, enhancing interaction effectiveness. ALGPT also builds specialized knowledge bases for each agent using CoT and In-Context Learning~\cite{brown2020language}.


\section{Datasets and Benchmark}
\label{sec:dataset}
We organize recent open-source work to foster research on advanced ADSs. Mainstream ADS datasets are summarized in Table~\ref{Table-dataset}.

\noindent \textbf{Single-Agent Autonomous Driving Data.} Single-agent datasets are obtained from a single reference agent, which can be the ego vehicle or roadside infrastructure, using various sensors.
Mainstream singel-agent autonomous driving datasets like KITTI~\cite{geiger2012we}, nuScenes~\cite{caesar2020nuscenes}, and Waymo~\cite{sun2020scalability} provide comprehensive multimodal sensor data, enabling researchers to develop and benchmark algorithms for multiple tasks such as object detection, object tracking, and object segmentation. In addition to these foundational datasets, newer ones like BDD-X~\cite{kim2018textual}, DriveLM~\cite{sima2025drivelm}, and nuScenes-QA~\cite{qian2024nuscenes} introduce action descriptions, detailed captions, and question-answer pairs that can be used to interact with LLMs. Combining language information with visual data can enrich semantic and contextual understanding, promote a deeper understanding of driving scenarios, and improve the safety and interaction capabilities of autonomous vehicles.

\noindent \textbf{Multi-agent Autonomous Driving Dataset.}
Beyond single-vehicle view datasets, integrating more viewpoints of traffic elements, such as drivers, vehicles and infrastructures into the data also brings advantages to AD systems. 
Multi-agent autonomous driving datasets, such as DAIR-V2X~\cite{yu2022dair}, V2XSet~\cite{xu2022v2x}, V2V4Real~\cite{xu2023v2v4real}, and TUMTraf-V2X~\cite{zimmer2024tumtraf}  typically include data from multiple vehicles or infrastructure sensors, capturing the interactions and dependencies between different agents and additional knowledge regarding the environments.
These datasets are essential for researching and developing cooperative perception, prediction, and planning strategies that enable vehicles to overcome the limitations of single agent datasets such as limited field of view (FOV) and occlusion.

\noindent \textbf{Benchmarks.} Several benchmarks are particularly well-suited for evaluating collaborative decision-making in autonomous driving. The INTERACTION dataset \citep{zhan2019interaction} includes a variety of real-world interactive scenarios, such as roundabouts and lane merging. It provides vehicle trajectories that enable an assessment of cooperative maneuvering and negotiation behaviors. Another important benchmark is the Waymo Open Motion Dataset \citep{ettinger2021large}, which is explicitly designed for interactive multi-agent motion prediction and planning. It features challenging scenarios, including merges and unprotected left turns, along with detailed annotations of interactive agents. In addition, the SMARTS benchmark \citep{zhou2021smarts} offers standardized scenarios for multi-agent autonomous driving research, particularly focusing on ramp merging and navigating unsignalized intersections. This work allows for direct comparisons of algorithms in cooperative traffic management tasks. These benchmarks provide comprehensive test bases for evaluating the coordination, safety, and adaptability of LLM-based multi-agent ADSs.

\section{Challenges and Future Directions}
\label{sec:opportunities}
This section explores key open challenges and potential opportunities for future research. 

\noindent \ding{182}~\textit{\textbf{Hallucination, Safety \& Trustworthiness.}}
Hallucination refers to LLMs generating outputs that are factually incorrect or non-sensical~\cite{huang2023survey}. 
In complex driving scenarios, a single driving agent's hallucinations in an LLM-based multi-agent ADS can be accepted and further propagated by other agents in the network via the inter-agent communication, potentially leading to serious accidents. 
Detecting agent-level hallucinations and managing inter-agent information flow are key to improving system safety and trust~\cite{fan2024hallucination}. Recent advances in spatiotemporal traffic analysis~\cite{zhang2024large,jiang2024empowering} further support real-time condition assessment, improving vehicle-road interaction and overall safety of ADS.

\noindent \ding{183}~\textit{\textbf {Legal, Security \& Privacy.}}
As agents autonomously exchange and process information within multi-agent ADS, the distribution of legal liability between individual users and manufacturers becomes ambiguous, particularly in cases involving system failures or collisions. In addition, vulnerable communication methods and strict user privacy requirements place high demands on cryptographic protocols and data management. These interrelated concerns collectively represent critical directions for future research and regulatory initiatives.

\noindent \ding{184}~\textit{\textbf {Multi-Modality Ability.}}
In current multi-agent systems, agents primarily use LLMs for scene understanding and decision-making. 
Perception outputs are converted into text via manual prompts or interpreters, then processed by LLMs to generate decisions. 
This pipeline is limited by perception performance and may cause information loss~\cite{gao2023llama}. Integrating language understanding with multimodal data fusion offers a promising direction for future multimodal multi-agent ADSs.

\noindent \ding{185}~\textit{\textbf{Real-World Deployment \& Scalability.}} 
LLM-based multi-agent ADS can scale up by adding more agents to handle increasingly complex driving scenarios. 
However, more LLM agents increase the demand for computing resources, while their interactions impose strict requirements on communication efficiency, which is critical for real-time decision-making~\cite{huang2024efficient}.
Therefore, under limited computing resources, it is crucial to develop a system architecture that supports distributed computing and efficient communication, as well as agents capable of adapting to various real-world environments and tasks, to optimize multi-agent ADS within resource constraints.

\noindent \ding{186}~\textit{\textbf{Human-Agent Interaction.}}
Current multi-agent ADS struggle to communicate intentions to human road users, relying on static signals inadequate for complex scenarios. Developing LLM-powered adaptive interfaces that generate context-appropriate, human-understandable communications while maintaining safety and trust presents a key deployment challenge~\cite{xia2025automating}.

\section{Conclusion}
\label{sec:conclusion}
This paper systematically reviews LLM-based multi-agent ADSs and traces their evolution from single-agent to multi-agent systems. 
We detail their core components, including agent environments and profiles, inter-agent interaction, and agent-human communication. 
Existing studies are categorized by interaction types and applications. 
We further compile public datasets and open-source implementations, and discuss challenges and future directions. 
We hope this review will inspire NLP community to explore more practical and impactful applications in LLM-based multi-agent ADS.

\section*{Limitations}
Despite being a survey, this work still has several limitations. \ding{182}~\textbf{Emerging Research and Limited Data.}  As LLM-based multi-agent ADS is an emerging field, the current body of research is still growing. While this may limit the breadth of our classification, we have aimed to provide a representative and forward-looking overview based on the most relevant and recent work.
\ding{183}~\textbf{Some Unverified Work.} Given the novelty of this topic, some referenced works are from unreviewed arXiv preprints. We include them to reflect the latest progress and ideas, while acknowledging that their findings may require further validation through peer review.
\ding{184}~\textbf{Limited Discussion on Real-world Applications.} Although industrial adoption of LLM-based multi-agent ADS is underway, public documentation remains limited. As a result, this review focuses on academic contributions, and real-world deployments are left for future investigation.

\bibliography{custom}

\appendix
\newpage

\section{Data-driven Autonomous Driving System}
\label{app:Data-driven}
Traditional ADS rely on data-driven approaches, which are categorized into modular and end-to-end frameworks~\cite{chen2024end}. 
\textbf{Modular-based systems} break the entire autonomous driving process into separate components, such as \textit{perception module}, \textit{prediction module}, and \textit{planning module}.
Perception modules are responsible for obtaining information about the vehicle's surrounding environment, aiming to identify and locate important traffic elements such as obstacles, pedestrians, and vehicles near the autonomous vehicle, usually including tasks such as object detection~\cite{wang2021detrd} and object occupancy prediction~\cite{tong2023scene}.
Prediction modules estimate the future motions of surrounding traffic participants based on the information provided by the perception module, usually including tasks such as trajectory prediction and motion prediction~\cite{shi2022motion}.
Planning module aims to derive safe and comfortable driving routes and decisions through the results of perception and prediction~\cite{sauer2018conditional}.
Each module is individually developed and integrated into onboard vehicles to achieve safe and efficient autonomous driving functions. 
Although modular methods have achieved remarkable results in many driving scenarios, the stacking design of multiple modules can lead to the loss of key information during transmission and introduce redundant calculations. 
Furthermore, due to the inconsistency in the optimization objectives of each module, the modular-based system may accumulate errors, which can negatively impact the vehicle's overall decision-making performance.
\textbf{End-to-end-based systems} integrate the entire driving process into a single neural network, and then directly optimize the entire driving pipeline from sensor inputs to produce driving actions~\cite{chen2024end}.
However, this method introduces the ``black box'' problem, meaning a lack of transparency in the decision-making process, complicating interpretation.


\section{LLMs in Autonomous Driving System}
\label{app:single}
\begin{figure*}[t]
\centering
     \includegraphics[width=0.92\textwidth]{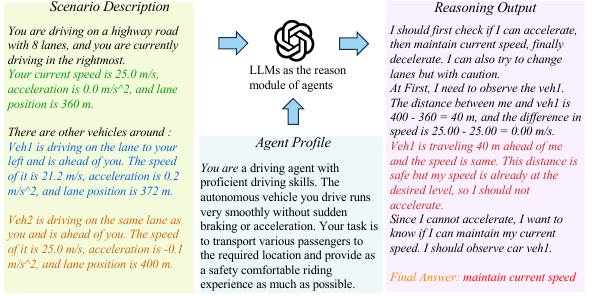}
    \caption{An example of an LLM-based single-agent ADS~\cite{wen2023dilu}.}
    \label{fig:single-agent}
\end{figure*}

\begin{figure*}[t]
\centering
     \includegraphics[width=0.92\textwidth]{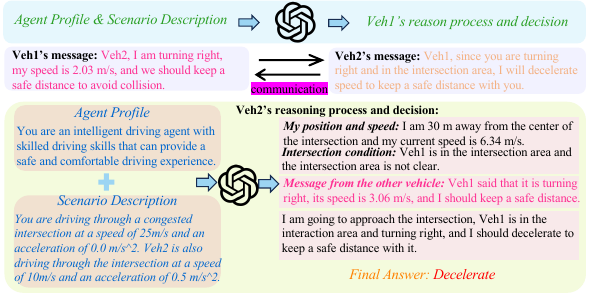}
    \caption{The communication among multiple agents in an LLM-based multi-agent system~\cite{hu2024agentscodriver}.}
    \label{fig:multi-agent}
\end{figure*}

As shown in Figure~\ref{fig:single-agent} and Figure~\ref{fig:multi-agent}, LLMs, with their powerful open-world cognitive and reasoning capabilities, have shown significant potential in ADSs~\cite{yang2023llm4drive,li2023towards}. 
LC-LLM~\cite{peng2024lc} is an explainable lane change prediction model that leverages LLMs to process driving scenario information as natural language prompts. By incorporating CoT reasoning and supervised finetuning, it not only predicts lane change intentions and trajectories but also provides transparent and reliable explanations for its predictions.
GPT-Driver~\cite{mao2023gpt} regards the motion planning task as a language modeling problem, using a fine-tuned GPT-3.5 model~\cite{ye2023comprehensive} to generate driving trajectories. 
DriveGPT4~\cite{xu2024drivegpt4} introduces an interpretable end-to-end autonomous driving system that uses multimodal LLMs to process multi-frame video inputs and textual queries, enabling vehicle action interpretation and low-level control prediction. By employing a visual instruction tuning dataset and mixfinetuning strategy, it provides a novel approach to directly map sensory inputs to actions, achieving superior performance in autonomous driving tasks.
Driving with LLM~\cite{chen2024driving} integrates vectorized numeric data with pre-trained LLMs to improve context understanding in driving scenarios and enhances the interpretability of driving decisions.

\section{LLM-enhanced Multi-Agent ADSs}
To highlight the application of LLMs and other NLP technologies in multi-agent ADSs, we have specially prepared Table~\ref{tab:llm_vehicle_sim}. This table systematically sorts out existing research from the two dimensions of ``environment \& subject characteristics'' and ``interaction mode'', and marks the LLMs used in each solution one by one.
Our goal is to help readers quickly grasp the landscape of this cross-domain research and better understand how LLM capabilities are being adapted to complex ADS scenarios.

\section{Real-World Multi-Agent LLM Systems in Autonomous Driving}
\noindent \textbf{NVIDIA DriveOS LLM Integration}\footnote{https://developer.nvidia.com/blog/streamline-llm-deployment-for-autonomous-vehicle-applications-with-nvidia-driveos-llm-sdk/} NVIDIA has released the DriveOS LLM SDK, which allows multiple AI agents (for perception, planning, user interaction) to run on edge computing, with the ability to infer LLMs. For example, a car can have a local LLM-based agent for real-time driving decisions or V2X message interpretation, offloading heavy computational tasks to optimized hardware. This lightweight onboard agent works in tandem with powerful cloud-based AI. In this conceptual multi-agent setup, the onboard LLM can handle immediate tasks and natural language commands while  querying the cloud-based LLM for complex planning or traffic knowledge.

\noindent \textbf{LLM in the Cabin and Beyond}\footnote{https://www.theverge.com/2023/6/16/23763208/mercedes-benz-chatgpt-voice-assistant-beta-test} Mercedes-Benz is collaborating with NVIDIA to develop multimodal LLMs that can interpret sensor data and driver preferences to assist in driving decisions. LLM-based intelligent agents can act as ``digital co-pilots'', monitoring the surrounding environment and providing maneuvering recommendations.

\renewcommand{\arraystretch}{1.4}

\definecolor{headercolor}{RGB}{70, 130, 180}   
\definecolor{modecolor}{RGB}{240, 240, 255}      
\definecolor{modesub}{RGB}{250, 250, 255}        
\definecolor{structurecolor}{RGB}{255, 240, 220} 
\definecolor{structuresub}{RGB}{255, 245, 230}   

\begin{table*}[!t]
\label{Tab:interaction}
\centering
\caption{Comparison of Interaction Modes and System Structures in LLM-Based Multi-Agent ADSs.}
\fontsize{8.5pt}{10pt}\selectfont
\begin{tabularx}{\textwidth}{%
>{\raggedright\arraybackslash}p{3.2cm}
>{\raggedright\arraybackslash}p{5.5cm}
>{\raggedright\arraybackslash}p{6cm}}
\toprule
\rowcolor{headercolor}\color{white}\textbf{Dimension} & \color{white}\textbf{Advantage} & \color{white}\textbf{Limitation} \\
\midrule

\rowcolor{modecolor}\textbf{Mode} & & \\
\rowcolor{modesub}\textit{Co-operative} & Enhances traffic flow efficiency and reduces collision risk by sharing agent intents and aligning actions. & Unexpected selfish behavior from uncooperative agents can propagate unsafe plans to the entire fleet. \\
\addlinespace[2pt]
\rowcolor{modesub}\textit{Competitive} & Can lead to more assertive and individually optimized behaviors in contested scenarios, such as securing a lane change in dense traffic. & Risks escalating conflicts and reducing overall traffic system stability if not properly regulated, potentially leading to gridlock or unsafe maneuvers. \\
\addlinespace[2pt]
\rowcolor{modesub}\textit{Debate} & LLM-based driving agents critique each other’s plans, surfacing hazards and converging on safer, more optimal strategies before execution. & Can lead to significant communication overhead and decision delay, which is a problem for real-time driving decisions. \\
\addlinespace[5pt]
\rowcolor{structurecolor}\textbf{Structure} & & \\
\rowcolor{structuresub}\textit{Centralised} & Enables strong global coordination and optimized system-wide decisions for traffic management due to a comprehensive overview. & Single-point failure and uplink delays can endanger all participating vehicles. \\
\addlinespace[2pt]
\rowcolor{structuresub}\textit{Decentralised} & Offers high robustness and scalability as individual agent failures have limited systemic impact, allowing for agile responses to local traffic conditions. & Lacks a global picture; local optima (e.g., platoon break-ups) can degrade overall traffic efficiency and safety. \\
\addlinespace[2pt]
\rowcolor{structuresub}\textit{Hierarchical} & Layered clusters (vehicle $\rightarrow$ platoon $\rightarrow$ cloud) scale to city-wide fleets while containing message volume within each tier. & Can introduce communication delays between layers and may suffer from inflexibility if the hierarchy is too rigid to adapt to highly dynamic situations. \\
\addlinespace[2pt]
\rowcolor{structuresub}\textit{Shared Message Pool} & Allows flexible, asynchronous information sharing, reducing direct communication burdens and enabling opportunistic coordination. & Contention and information overload risk stale or conflicting data, demanding strict access control. \\

\bottomrule
\end{tabularx}
\label{tab:driving_dimensions}
\end{table*}

\newcolumntype{C}[1]{>{\centering\arraybackslash}p{#1}}
\newcolumntype{I}[1]{>{\itshape\centering\arraybackslash}p{#1}}
\newcolumntype{L}[1]{>{\raggedright\arraybackslash}p{#1}}

\definecolor{HumanClr}{RGB}{230,210,255}
\definecolor{interactClr}{RGB}{190,215,255}
\definecolor{communClr}{RGB}{206,240,220}
\definecolor{orchestClr}{RGB}{255,196,141}  
\definecolor{headerClr}{RGB}{150,180,240}    
\colorlet{interactLight}{interactClr!55}
\colorlet{orchestLight}{orchestClr!55}
\colorlet{communLight}{communClr!55}
\colorlet{HumanLight}{HumanClr!55}

\definecolor{rowA}{RGB}{248,248,248}
\rowcolors{3}{rowA}{white}

\setlength{\tabcolsep}{0.1pt}
\renewcommand{\arraystretch}{1.6}

\begin{table*}[htbp]
  \centering
  \captionsetup{
    width=18cm,
    justification=centering,
    singlelinecheck=true
  }
  \caption{Comparative Summary of LLM-Based Multi-Agent ADS Research.}
  \label{tab:llm_vehicle_sim}

  \hspace*{-1cm}
  {\fontsize{6.5pt}{7.5pt}\selectfont
  \begin{tabularx}{15cm}{@{}L{2.0cm} C{1cm} C{1.7cm} C{1.7cm} C{1.5cm} C{1.5cm} C{1.5cm} C{2.0cm} C{2.0cm} C{2.0cm}@{}}
    \toprule
    \rowcolor{headerClr}
    \textbf{Paper} & \textbf{Date} & \textbf{Environment} & \textbf{Profile-Method} & \textbf{Profile-Setting} & \textbf{Structure} & \textbf{Mode} & \textbf{Human-Feedback} & \textbf{LLM Model} \\
    \midrule
    LanguageMPC\newline\citep{sha2023languagempc}              & 2023/10 & Simulation & Pre-defined & Vehicle agents, Human & Centralized   & Cooperative & Instructor Paradigm                    & GPT-3.5   \\
    AgentsCoDriver\newline\citep{hu2024agentscodriver} & 2024/04          & Simulation & Pre-defined & Vehicle agents & Decentralized & Cooperative & -                   & GPT-3.5-turbo      \\
    KoMA\newline\citep{jiang2024koma}                & 2024/07 & Simulation & Pre-defined & Vehicle agents & Shared Message pool   & Cooperative & Instructor Paradigm                         & GPT-4               \\
    AgentsCoMerge\newline\citep{hu2024agentscomerge}     & 2024/08       & Simulation & Pre-defined & Vehicle agents & Decentralized, Hierarchical & Cooperative & Instructor Paradigm       & GPT/Claude/Gemini Series               \\
    CoDrivingLLM\newline\citep{fang2024towards}       & 2024/09      & Simulation & Pre-defined & Vehicle agents & Centralized   & Cooperative & Instructor Paradigm      & GPT-4o               \\
    CoMAL\newline\citep{yao2024comal}      &  2024/10           & Simulation & Pre-defined & Vehicle agents, Human & Shared Message pool & Cooperative & Instructor Paradigm                        & GPT-4o-mini, Qwen-72B, Qwen-32B, Qwen-7B               \\
    Complement-Vehicle’s-FOV\newline\citep{dona2024tapping} & 2024/08 & Simulation & Pre-defined & Vehicle agents, Infrastructure agents, Human & Decentralized, Hierarchical, Centralized    & Cooperative & Instructor Paradigm, Partnership Paradigm       & GPT-4V, GPT-4o               \\
    CAV-LLM-Driving-Assistant\newline\citep{tang2024test} & 2024/11 & Simulation & Pre-defined & Vehicle agents,Human & Decentralized & Cooperative & Instructor Paradigm                        & GPT-4V, GPT-4o               \\
    EC-Drive\newline\citep{chen2024edge}     &    2024/08         & Simulation & Pre-defined & Vehicle agents, Infrastructure agents & Hierarchical   & Cooperative & Instructor Paradigm   & LLaMA-Adapter (7B), GPT-4              \\
    ChatSim\newline\citep{wei2024editable}     &   2024/02         & Simulation & Pre-defined, Model-generated & Human, Assistant agents & Hierarchical, Centralized & Cooperative & Instructor Paradigm          & GPT-4              \\
    ALGPT\newline\citep{zhou2024algpt}    &     2024/01 &       Simulation & Pre-defined, Model-generated & Assistant agents & Hierarchical   & Cooperative & -                        & GPT series              \\
    AD-H\newline\citep{zhang2024ad}          &2024/06           & Simulation & Pre-defined & Vehicle agents, Human & Hierarchical & Cooperative & Instructor Paradigm  & LLaVA-7B-V1.5              \\
    SurrealDriver\newline\citep{jin2024surrealdriver} & 2023/09 & Simulation & Pre-defined & Vehicle agents, Infrastructure agents, Human & Hierarchical   & Cooperative & Instructor Paradigm                          & GPT series, Llama, PaLM              \\
    LDPD\newline\citep{liu2024language}              & 2024/10       & Simulation & Model-generated & Vehicle agents & Hierarchical, Centralized & Cooperative & -      & GPT-4o              \\
    V-HOI MLCR\newline\citep{zhang2024enhancing}         & 2024/03      & Simulation & Pre-defined & Vehicle agents, Human & Hierarchical   & Cooperative, Debate & Instructor Paradigm                & GPT-4, GPT-3.5              \\
    Co-Pilot\newline\citep{wang2023chatgpt}         & 2023        & Physics & Pre-defined & Vehicle agents, Human & Decentralized & Cooperative & Instructor Paradigm                         & GPT-3.5-turbo-0301              \\
    PPE\newline\citep{ma2024learning}              & 2024        & Simulation & Pre-defined & Vehicle agents, Human & Decentralized   & Cooperative & Partnership Paradigm                         & GPT-4-turbo-preview and GPT-3.5-turbo              \\
    Drive-as-You-Speak\newline\citep{cui2024drive}   & 2023/09   & Simulation & Pre-defined & Vehicle agents, Human & Decentralized & Cooperative & Instructor Paradigm, Partnership Paradigm       & GPT-4              \\
    Reason-and-React\newline\citep{cui2024receive}         & 2023/10 & Simulation & Pre-defined & Vehicle agents, Human & Decentralized   & Cooperative & Instructor Paradigm, Partnership Paradigm    & GPT-4              \\
    DriVLMe\newline\citep{huang2024drivlme}           & 2024/06       & Simulation & Pre-defined & Vehicle agents, Human & Decentralized & Cooperative & Instructor Paradigm          & Vicuna-7B + LoRA              \\
    AccidentGPT\newline\citep{wang2024accidentgpt}      & 2024/06        & Physics & Pre-defined & Vehicle agents, Infrastructure agents, Human & Hierarchical, Centralized, Decentralized   & Cooperative & Instructor Paradigm   & GPT-4              \\
    ConnectGPT\newline\citep{tong2024connectgpt}     & 2024/06          & Physics & Pre-defined & Vehicle agents, Infrastructure agents, Human & Hierarchical, Centralized, Decentralized & Cooperative & Instructor Paradigm  & GPT-4              \\
    DriveAgent\newline\citep{hou2025driveagent}     & 2025/05         & Physics & Pre-defined & Assistant agents & Decentralized   & Cooperative & -  & LLaMA-3.2-Vision              \\
    CCMA\newline\citep{zhang2025ccma}               & 2025      & Simulation & Pre-defined & Vehicle agents, Assistant agents & Hierarchical, Decentralized & Cooperative & -     & GLM-4v-9B              \\
    V2V-LLM\newline\citep{chiu2025v2v}             & 2025/02     & Simulation & Pre-defined & Vehicle agents & Decentralized   & Cooperative & - & LLaVA-v1.5-7b              \\
    IITI\newline\citep{fang2025interact}         & 2025/03  & Simulation & Pre-defined & Vehicle agents, Human & Decentralized & Cooperative & Instructor Paradigm     & Llama3              \\
    Tell-drive\newline\citep{xu2025tell}         & 2025/02  & Simulation & Pre-defined & Vehicle agents & Hierarchical, Decentralized & Cooperative & -    &  GPT-4o-min              \\
    Human-RLHF\newline\citep{sun2024optimizing}         & 2024/06  & Simulation & Pre-defined & Vehicle agents, Human & Decentralized & Cooperative & Instructor Paradigm          & GPT-4o              \\
    GameChat\newline\citep{mahadevan2025gamechat}   & 2025/03        & Simulation & Pre-defined & Vehicle agents, Human & Decentralized & Cooperative & Instructor Paradigm       & GPT-4o-mini               \\
    hybrid LLM-DDQN\newline\citep{yan2025hybrid}  & 2024/10         & Simulation & Pre-defined & Vehicle agents, Infrastructure agents & Decentralized, Hierarchical & Cooperative & -     & GPT-3.5, Llama3.1-8B, Llama3.1-70B              \\
    \bottomrule
  \end{tabularx}}
  \hspace*{-1.2cm}
\end{table*}

\end{document}